\documentclass[aps,twocolumn,pra,showpacs,floatfix,superscriptaddress]{revtex4-1}
\usepackage{epsfig}
\usepackage{graphicx}
\usepackage{dcolumn}
\usepackage{amssymb,amsmath}
\usepackage{mathrsfs}
\usepackage{nicefrac}

\newcommand{\vect}[1]{\ensuremath{\mathbf{#1}}}

\newcommand{\ket}[1]{\ensuremath{\left| #1 \right>}}
\newcommand{\rme}[3]{\ensuremath{\langle #1 || #2 || #3 \rangle}}

\newcommand{\half}{\nicefrac{1}{2}}

\begin{document}

\title{Screening of oscillating external electric field in atoms}

\author{V. A. Dzuba}
\affiliation{School of Physics, University of New South Wales, Sydney, NSW 2052, Australia}
\author{J. C. Berengut}
\affiliation{School of Physics, University of New South Wales, Sydney, NSW 2052, Australia}
\author{J. S. M. Ginges}
\affiliation{School of Mathematics and Physics, The University of Queensland, Brisbane QLD 4072, Australia}
\author{V. V. Flambaum}
\affiliation{School of Physics, University of New South Wales, Sydney, NSW 2052, Australia}

\begin{abstract}
We study the screening of a homogeneous oscillating external electric field $E_0$ in noble-gas atoms using atomic many-body calculations.
At  zero frequency  of the oscillations ($\omega=0$) the screened field $E(r)$ vanishes at the nucleus, $E(0)=0$. However, the profile of the field $E(r)$ is complicated, with the magnitude of the field
exceeding the external field $E_0$ at certain points. For $\omega >0$ the field $E(r,\omega)$ strongly depends on $\omega$ and at some points may exceed the external field $E_0$  many times. The field at the nucleus is not totally screened and grows with $\omega$ faster than $\omega^2$. It  can even be enhanced when $\omega$ comes close to resonance with a frequency of an atomic transition. This field interacts with CP-violating nuclear electric dipole moments creating new opportunities for studying them. The screening of the external field by atomic electrons may strongly suppress (or enhance near an atomic resonance) the low energy nuclear electric dipole transitions.
\end{abstract}


\maketitle

\section{Introduction}

Nuclear forces that violate the conservation of combined charge-conjugation and parity (CP) produce CP-violating nuclear moments that in turn may
produce observable effects in atoms and molecules. The study of these effects provides a powerful probe of new physics
beyond the Standard Model; see, e.g., reviews \cite{FGreview,PRreview,ERKreview,Sreview,CFRSreview}. According to the Schiff theorem~\cite{Schiff}, the lowest-order CP-violating  moment, 
the nuclear electric dipole moment (EDM), is unobservable in neutral atoms. Indeed, a neutral atom (and its nucleus) is not accelerated in a
homogeneous static external electric field. Considering the nucleus to
be point-like, the external electric field is completely screened at the nucleus by atomic electrons, and the nuclear EDM has nothing to interact with. 

This screening is a big obstacle in the study of CP-violating nuclear forces. One has to go to higher-order moments or include 
some small corrections. For example, the screening is not complete if finite nuclear size is taken into 
account~\cite{Schiff}. Indeed, while the total force on the nucleus is zero, %
the electric field does not have to vanish at each point across the nucleus.  A convenient way to consider this effect is by introducing the so-called nuclear
Schiff moment which induces atomic and molecular EDMs~\cite{Sandars,Hinds,SFK,FKS1986}. It can be roughly described as what is left from the nuclear EDM when  
the screening of the external electric field by electrons is taken into account. Refs. ~\cite{Sandars,Hinds} considered the effect of the
proton EDM. The Schiff moments produced by CP-violating nuclear forces
were introduced and calculated for the first time in
~\cite{SFK,FKS1986}.  Among other possibilities is the atomic EDM generated by the nuclear magnetic quadrupole moment ~\cite{SFK}.

The electric field is not totally screened in ions. However, the strong constant electric field would remove ions from the trap. 
Another possibility is to use an oscillating electric field. It was 
stated in Ref.~\cite{DFSS86} that an oscillating electric field is not totally screened in atoms and may even
lead to an enhancement of the field at the nucleus when the frequency of the field oscillations is close 
to the frequency of an atomic transition. In a recent work~\cite{Flambaum18}, a formula was
derived which states that the screened oscillating field at the nucleus
 is proportional to $\omega^2\alpha_{zz}(\omega)$, where $\omega$ is the frequency of the
oscillating field and $\alpha_{zz}(\omega)$ is the dynamic polarizability of the atom at this frequency. 
At sufficiently large frequencies the screening is significantly reduced. The field at the nucleus may  
even be enhanced in the resonance situation when the frequency of the external field is close to the frequency 
of an atomic transition.

In this paper we study the effect of screening of the external oscillating electric field in the noble-gas atoms numerically using
the relativistic time-dependent Hartree-Fock method which is also known as the random-phase approximation. We demonstrate that the numerical calculations agree practically exactly with the 
formula for the screened electric field at the nucleus from Ref.~\cite{Flambaum18}.
Thus we have checked that  
the problem of finding the screened oscillating field in atoms is reduced to the calculations (or measurements) of
the atomic dynamic polarizabilities. This in turn may lead to new ways of studying nuclear EDMs. Another application is the calculation of the effect of the electron screening on the probabilities of the nuclear electric dipole transitions. 
  
In this paper we have also calculated the screened electric field inside an atom at all distances. The screened field oscillates and can actually exceed the external field.

\section{Calculations}
It has been shown in Ref.~\cite{Flambaum18} that an external
oscillating electric field is screened at the atomic
nucleus with the value
 \begin{equation}
E= \frac{E_0}{Z}
\left[ Z_i - {\tilde \omega}^2 {\bf \tilde  \alpha}_{zz}\right],
\label{eq:E0}
\end{equation}
where $E_0$ is the amplitude of an external field directed along the $z$ axis, $Z$ is the nuclear charge, $Z_i$ is the ionization
degree, ${\tilde \omega} = \frac{\omega}{e^2/\hbar a_b}$ is the
oscillation frequency $\omega$ in atomic units, 
${\tilde  \alpha}_{zz}=\frac{ \alpha_{zz}(\omega)}{a_b^3}$ is the
dynamic polarizability of the atom $\alpha_{zz}(\omega)$ in atomic
units, and $a_b$ is the Bohr radius. 
For $Z_i=0$ and  $\omega=0$, the field at the nucleus is totally screened, $E=0$, in agreement with the Schiff theorem.

It is known that the Schiff theorem is fulfilled exactly in the random-phase approximation (RPA)~\cite{DFSS86}.
RPA can be considered as a self-consistent Hartree-Fock approximation
in a weak external field so that only terms linear in 
the external field are kept. It is also known that the RPA
approximation gives very accurate values for the atomic 
polarizabilities for noble-gas atoms (see, e.g., Ref.~\cite{Safronova} and below). This means that the use of the
RPA method for noble-gas atoms is a good starting point for studying the screening of the external electric
field in atoms. Note that for the closed-shell atoms the $\alpha_{zz}$
polarizability in Eq.~(\ref{eq:E0}) is just the scalar polarizability $\alpha_0$.

\subsection{Random-phase approximation}

We start from the Hartree-Fock equations for the single-electron orbital $\psi_a$ (atomic units, $\hbar = m_e = |e| = 1$):
\begin{gather}
\label{eq:HF} 
(\hat H_0 -\epsilon_a) \psi_a = 0, \\
\hat H_0 = c\,\boldsymbol{\alpha}\cdot\vect{p} + (\beta-1)\,c^2
+V_{\rm nuc}+ \hat V. \nonumber
\end{gather}
$\hat H_0$ is the relativistic Hartree-Fock Hamiltonian, 
$V_{\rm nuc}\approx -Z/r$ is the finite-size nuclear potential, 
$\psi_a(\vect{r})$ is a four-component Dirac spinor,
\[
\psi_a(\vect{r}) = \frac{1}{r}\left( f(r)\,\Omega_{\kappa m} \atop ig(r)\,\Omega_{-\kappa m} \right)
\]
and $\hat V$ is the self-consistent electronic potential
\begin{multline}
\hat V \psi_a(\vect{r}) = \sum_b \int d^3r' \frac{\psi_b^\dag(\vect{r}') \psi_b(\vect{r}')}{|\vect{r}-\vect{r'}|} \psi_a(\vect{r}) \\
- \sum_b \int d^3r' \frac{\psi_b^\dag(\vect{r}') \psi_a(\vect{r}')}{|\vect{r}-\vect{r'}|} \psi_b(\vect{r})
\end{multline}
where the index $b$ enumerates the electrons in the core.
An applied weak periodic field
\begin{equation}
\label{eq:F} 
\hat F = \hat f^{-i\omega t} + \hat f^{\dagger} e^{i\omega t},
\end{equation}
modifies the atomic orbitals, adding to them small oscillating corrections
\begin{equation}
\label{eq:psi} 
\tilde \psi_b = \psi_b + \chi_b e^{-i\omega t} + \eta_b e^{i\omega t},
\end{equation}
which can be found by solving the RPA equations
\begin{eqnarray}
&& (\hat H_0 -\epsilon_b - \omega) \chi_b = - (\hat f +\delta \hat V)\psi_b, \nonumber \\
&& (\hat H_0 -\epsilon_b + \omega) \eta_b = - (\hat f^{\dagger}
   +\delta \hat V^{\dagger})\psi_b, \label{eq:RPA}
\end{eqnarray}
where $\delta \hat V$ is the correction to the self-consistent Hartree-Fock potential
due to the external field. We consider the case where $\hat f$ is the electric dipole operator 
(in length form $\hat f =z$).
Equations (\ref{eq:RPA}) are solved self-consistently for all states in the core.

Detailed equations for (\ref{eq:RPA}) can be found in~\cite{johnson80pscr,DFS84}. Briefly, we expand the $\chi_b$ and $\eta_b$ in partial waves ($\chi_\beta$ and $\eta_\beta$) with fixed angular momentum $j_\beta$ and parity \mbox{$\pi = (-1)^{l_b+K}$} for electric $2^K$-pole excitations where $K$ is the rank of $\hat f$ ($K=1$ for the electric dipole operator) and $|j_b - K| \leq j_\beta \leq j_b + K$. The reduced matrix elements required are
\begin{subequations}
	\label{eq:RPAME}
\begin{align}
\rme{\chi_\alpha}{\delta \hat V&}{\psi_a} \nonumber \\
=~\sum_{b\beta}& \frac{\rme{\kappa_\alpha}{C^K}{\kappa_a}\rme{\kappa_\beta}{C^K}{\kappa_b}}{2K+1} 
	\label{eq:RPAME_dir} \\
 & \left( R^K(\chi_\alpha \psi_b,\psi_a\chi_\beta)+R^K(\chi_\alpha\psi_b,\psi_a\eta_\beta)\right) \nonumber \\
+\sum_{b \beta k}& (-1)^{K+k}\rme{\kappa_\beta}{C^k}{\kappa_a}\rme{\kappa_b}{C^k}{\kappa_\alpha} \nonumber \\
 & \left\{\begin{array}{ccc} j_a & j_\beta & k \\ j_b & j_\alpha & K \end{array}\right\}
   R^k(\chi_\alpha\psi_a,\psi_b\eta_\beta) \\
+\sum_{b \beta k}& (-1)^{K+k}\rme{\kappa_b}{C^k}{\kappa_a}\rme{\kappa_\alpha}{C^k}{\kappa_\beta} \nonumber \\
 & \left\{\begin{array}{ccc} j_a & j_b & k \\ j_\beta & j_\alpha & K \end{array}\right\}
 R^k(\chi_\alpha\psi_a,\chi_\beta\psi_b)
\end{align}
\end{subequations}
where $b$ runs over core states and the $\beta$ enumerate the partial waves of their respective corrections.
The reduced spherical tensor matrix elements are defined by
\begin{multline*}
\rme{\kappa_a}{C^k}{\kappa_b} = (-1)^{j_a + \half} \sqrt{(2j_a+1)(2j_b+1)}\\
\left( \begin{array}{ccc} j_a & j_b & k \\ -\half & \half & 0 \end{array} \right)
\xi(l_a + l_b + k)
\end{multline*}
with $\xi(n) = [(-1)^n + 1]/2$, while the radial Slater integrals are defined
\begin{gather*}
R^k(\psi_a\psi_b,\psi_c\psi_d) = \int dr\, \big(f_a(r)f_c(r)+g_a(r)g_c(r)\big) Y^k_{\psi_b \psi_d}(r) \\
Y^k_{\psi_b \psi_d}(r) = \int dr' \frac{r_<^k}{r_>^{k+1}}
\big(f_b(r')f_d(r')+g_b(r')g_d(r')\big),
\end{gather*}
where $r_< = \min(r,r')$ and $r_> = \max(r,r')$.

The conjugate equations $\rme{\eta_\alpha}{\delta \hat V^\dag}{\psi_a}$ are similar to (\ref{eq:RPAME}) but with $\chi_\alpha\rightarrow\eta_\alpha$ and $\chi_\beta \leftrightarrow \eta_\beta$ exchanged. In this work we are interested in the electric dipole polarizability, hence $K = 1$.

\subsection{Scalar polarizability}

The dynamic scalar polarizability of a closed-shell atom in the RPA method is given by
\begin{equation}
\label{eq:alphaw} 
\alpha_0(\omega) = -\frac{1}{3} \sum_{b\beta} 
\left(\langle \psi_b||\hat f|| \chi_\beta \rangle+
\langle \psi_b||\hat f|| \eta_\beta \rangle \right)
\end{equation}
where $b$ runs over core states.
Note that one can use summation over a complete set of the single-electron basis states $\ket{n}$ 
to calculate the corrections $\chi_b$ and $\eta_b$
\begin{eqnarray}
&& \chi_b = \sum_n \frac{\rme{n}{\hat f + \delta \hat V}{b}}{\epsilon_b - \epsilon_n + \omega}
|n\rangle, \label{eq:chi} \\
&& \eta_b = \sum_n \frac{\rme{n}{\hat f^\dagger + \delta \hat V^\dagger}{b}}{\epsilon_b - \epsilon_n - \omega}
|n\rangle \label{eq:eta}.
\end{eqnarray}
This would lead to a more commonly used expression for the dynamic polarizability of a closed-shell atom,
\begin{equation}
\alpha_0 = -\frac{2}{3}\sum_{bn} \frac{(\epsilon_b-\epsilon_n)
\langle b ||\hat f||n\rangle \langle n ||\hat f + \delta \hat V||b\rangle}{(\epsilon_b - \epsilon_n)^2 - \omega^2}.
\label{eq:asum}
\end{equation}
Summation in (\ref{eq:asum}) goes over occupied single-electron states $b$ and vacant states $n$.
We do not use expressions (\ref{eq:chi}), (\ref{eq:eta}) and (\ref{eq:asum}) in the calculations of the present work.
However, having these expressions is useful for a discussion of the polarizability behaviour near a resonance
($\omega \approx \epsilon_n - \epsilon_b$).

The induced electric potential inside the atom can be extracted from the direct term of (\ref{eq:RPAME_dir}) as
\begin{equation}
\delta V(r) = \frac{1}{3}\sum_{b\beta} \rme{\kappa_\beta}{C^1}{\kappa_b} 
\left( Y^1_{\psi_b \chi_\beta}(r) + Y^1_{\psi_b \eta_\beta}(r) \right),
\end{equation}
and the total screened electric field inside the atom is given by
\begin{equation}
\label{eq:e0}
E(r) = E_0 + \varepsilon(r) = E_0\big(1 +\frac{d}{dr}\delta V(r)\big)\ .
\end{equation}
Note that the derivatives of $Y^1_{\psi_b\psi_d}(r)$ can be expressed
\begin{multline*}
\frac{d}{dr} Y^1_{\psi_b\psi_d}(r)
= -\frac{2}{r^3}\int_0^r r' \big( f_b(r')f_d(r')+g_b(r')g_d(r')\big) dr' \\
  + \int_r^\infty \big( f_b(r')f_d(r')+g_b(r')g_d(r')\big)/r'^2\,dr'.
\end{multline*}

\begin{table}
\caption{\label{t:a0}
Static dipole polarizabilities of the noble-gas atoms calculated using the RPA method, experimental values 
for the polarizabilities, and the experimental positions of the first excitation which gives the dominant contribution
to the polarizability.}
\begin{ruledtabular}
\begin{tabular}{cccc}
\multicolumn{1}{c}{Atom}&
\multicolumn{2}{c}{$\alpha_0$ (a.u.)}&
\multicolumn{1}{c}{$\hbar \omega$ (cm$^{-1}$)}\\
\cline{2-3}
&\multicolumn{1}{c}{RPA}&
\multicolumn{1}{c}{Expt.}& \multicolumn{1}{c}{\cite{NIST}}\\
\hline
He & 1.322 & 1.383759\,(13)~\cite{He} & 169087 \\
Ne & 2.380 &  2.66110\,(3)~\cite{Ne} & 134042 \\
Ar  & 10.77  & 11.083\,(2)~\cite{Ar} & 93750 \\
Kr &  16.47 & 16.740~\cite{Kr} & 80917 \\
Xe &  26.97 & 27.292~\cite{Xe} & 68045 \\
Rn &  35.00 &                           & 55989 \\
\end{tabular}
\end{ruledtabular}
\end{table}

\begin{figure}[tb]
\epsfig{figure=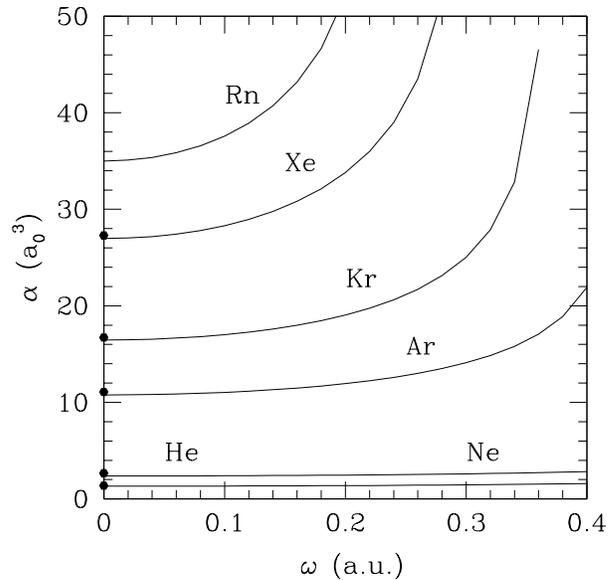,scale=0.4}
\caption{Dynamic plarizabilities of noble-gas atoms calculated in the RPA approximation.
Dots at $\omega=0$ show experimental values for static polarizabilities.}
\label{f:polw}
\end{figure}

Static dipole polarizabilities of the noble-gas atoms calculated using the RPA equations (\ref{eq:RPA}) and 
(\ref{eq:alphaw}) at $\omega=0$ are presented in Table~\ref{t:a0} and compared to the most accurate
experimental values. The difference is only a few per cent and tends to be better for the heavier atoms. 
There is no experimental value for Rn, however our calculated value 35.00\,a.u. agrees within 5.5\% with the
value 33.18\,a.u. obtained in the more sophisticated coupled-cluster calculations of Ref.~\cite{Rn}.
Table~\ref{t:a0} also presents experimental energies of the first excitation from the ground state which gives the 
dominant contribution to the polarizability at small frequency. These energies decrease in value monotonically
from He to Rn. This explains the larger polarizabilities for the heavier atoms and their faster increase with $\omega$
(see Fig.~\ref{f:polw}). Fig.~\ref{f:polw} shows dynamic polarizabilities of the noble-gas atoms calculated in
the RPA method using Eqs.~(\ref{eq:RPA}) and (\ref{eq:alphaw}). 

\begin{figure}[tb]
\epsfig{figure=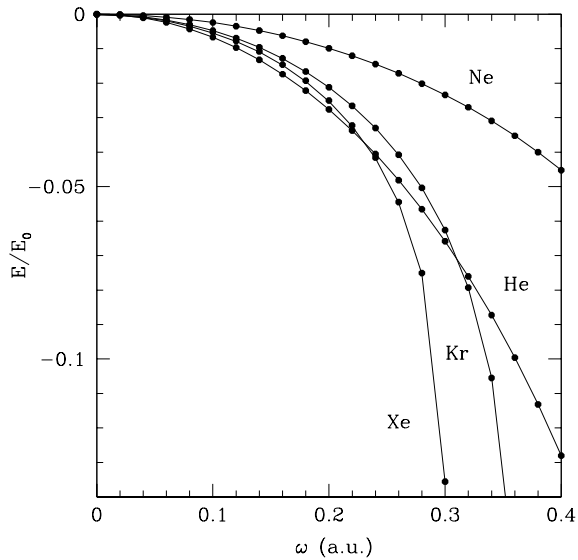,scale=0.4}
\caption{Screened electric field at the nuclei of He, Ne, Kr and Xe as a function of the frequency of 
the field oscillations. Solid line shows the result of the RPA calculations using formula (\ref{eq:e0}), dots come 
from formula (\ref{eq:E0}) with the calculated dynamic polarizability as in Fig.~\ref{f:polw}. 
The graphs for Ar and Rn are not shown because they are similar to those for Kr and Xe, respectively.}
\label{f:e0}
\end{figure}

Fig.~\ref{f:e0} shows the screened electric field on the nuclei as a function of the frequency of the field 
oscillations. Note the excellent agreement between the two methods of calculation, using (\ref{eq:E0}) or (\ref{eq:e0}).  
For $\omega <0.24$ a.u. the largest electric field on the nucleus is in the He atom, for $\omega >0.24$ a.u. 
the largest field is in Xe and Rn. For most of the noble-gas atoms (with an exception of Ne) the screening
is less than a tenth for $\omega > 0.3$ a.u. ($\lambda < 152$ nm ). Note that the field can even be 
enhanced~\cite{Flambaum18} when its frequency comes close to a resonance with an atomic transition.
The numerical method used in this paper does not allow us to come close to a resonance. Therefore, we leave this for a future study.

\begin{figure}[tb]
\epsfig{figure=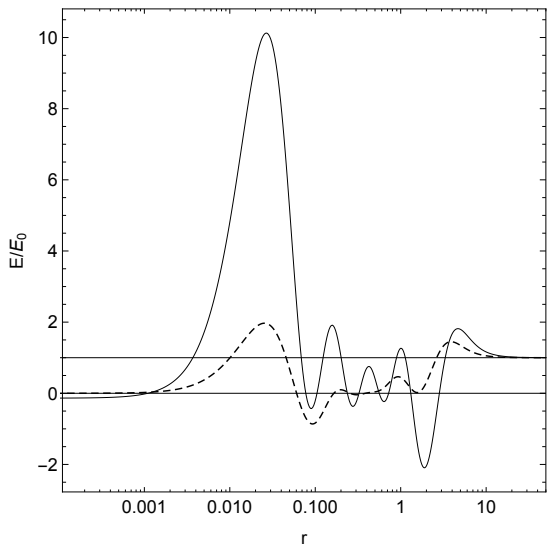,width=0.4\textwidth}
\caption{Screened external electric field in Xe as a function of the distance ($r$ is in atomic units).  
Dotted line corresponds to $\omega =0$, solid line is for $\omega =0.3$ a.u.}
\label{f:xe}
\end{figure}

\begin{figure}[tb]
\epsfig{figure=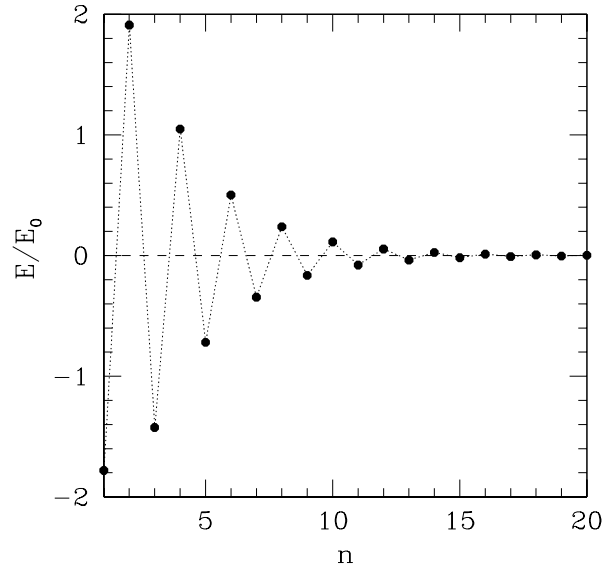,scale=0.4}
\caption{Calculated screened external electric field at $r=0$ in Xe as a function of the iteration number. }
\label{f:xe0}
\end{figure}

In contrast to the formula (\ref{eq:E0}) which gives the screened electric field at one point, $r=0$, the formula
 (\ref{eq:e0}) gives the screened electric field at any distance from the nucleus. Fig.~\ref{f:xe} shows the
 screened electric field 
 in Xe at two values of the frequency, $\omega=0$ and $\omega=0.3$ a.u. The field $E=E_0$ at
 large distances and its screened value at short distances is equal to what is given by formula (\ref{eq:E0}).
 However, inside the atom the behaviour is very complicated, reflecting the shell structure
 of the atom and oscillations of the wave functions of external 
 electrons. Note the strong enhancement of the peaks at $\omega>0$. This complicated behaviour 
 is a collective effect caused by the fine tuning of electron orbitals
 affected by the external field and the change in other orbitals. The collectiveness of the effect is
 illustrated in Fig.~\ref{f:xe0}. It shows the 
 electric field at the nucleus of the Xe atom at $\omega=0$ as a function of the iteration number. The iterations 
 are used to solve the RPA equations (\ref{eq:RPA}) starting from $\chi=0$ and $\eta=0$. Each iteration
 corresponds to the next order of perturbation theory in the residual Coulomb interaction. It takes about
 twenty iterations to get the correct field at the nucleus. This illustrates that the effect is not perturbative and has
 a collective nature. Similar pictures for Tl$^+$ were presented in our earlier work~\cite{DFSS86}. Note
 that the figure captions were misplaced in that work. 

For atoms other than the noble-gas atoms the correlations between external electrons and between external electrons
and the core electrons play an important role (see, e.g.,~\cite{Safronova,CI+MBPT}). These correlations are not included
in the RPA calculations. This means that neither formula (\ref{eq:alphaw}) for the dynamic polarizability nor
formula (\ref{eq:e0}) for the screened electric field are likely to give accurate results. However, the formula
(\ref{eq:E0}) was obtained without any assumptions about electron structure and should work well for
any atom. This reduces the problem of screening to the problem of the dynamic polarizability of an atom
which can be found from calculations or measurements.

For atoms with total angular momentum $J>1$ in the ground state the scalar polarizability $\alpha_0(\omega)$ 
should be replaced by $\alpha_{zz}(\omega)$ (see Eq. (\ref{eq:E0})) which may have vector and tensor contributions.
Calculation of the polarizabilities can be performed to very high accuracy for atoms with few
valence electrons above closed shells (see, e.g., \cite{Safronova,DD10}). However, even for atoms with more complicated
electron structure, e.g. for atoms with an open $f$-shell, the polarizabilities can still be calculated
with reasonably good accuracy ~\cite{DFL11,DKF14}.

\acknowledgments

This work was funded in part by the Australian Research Council
through a Discovery Project DP150101405 and a Future Fellowship (J.G.) FT170100452 
and by the National Science Foundation under grant No. NSF PHY11-25915. V.F. is grateful to Kavli Institute for Theoretical Physics at Santa Barbara for hospitality.

\end{document}